\documentclass[a4paper]{spie}  
\usepackage[]{graphicx,journals}

\newcommand\op{\frac{1}{12}}
\newcommand\oa{\frac{1}{6}}
\newcommand\ob{\frac{1}{6}e^{i\pi}}
\newcommand\oc{\frac{1}{6}e^{i\pi/2}}
\newcommand\od{\frac{1}{6}e^{i3\pi/2}}

\title{
Characterization of integrated optics components for the second
generation of VLTI instruments } 

\author{S.~Lacour\supit{a} L.~Jocou\supit{a} T.~Moulin\supit{a}
P.\,R.~Labeye\supit{b}
M.~Benisty\supit{a} J.-P.~Berger\supit{a} A.~Delboulb\'e\supit{a}
X.~Haubois\supit{c} E.~Herwats\supit{a} P.\,Y.~Kern\supit{a}
F.~Malbet\supit{a} K.~Rousselet-Perraut\supit{a} G.~Perrin\supit{c}
\skiplinehalf
\supit{a} LAOG, BP 53, 38041 Grenoble, France \\
\supit{b} CEA-LETI, Minatec,  17, Rue des Martyrs, 38054 Grenoble, France \\
\supit{c} LESIA, Observatoire de Paris/Meudon, 7 place Jules Janssen, 92190 Meudon, France
}

\begin{document} 
  \maketitle 

\begin{abstract}
Two of the three instruments proposed to ESO for the second generation
instrumentation of the VLTI would use integrated optics for beam
combination. Several design are studied, including co-axial and
multi-axial recombination. An extensive quantity of combiners are
therefore under test in our laboratories. We will present the various
components, and the method used to validate and compare the different
combiners. Finally, we will discuss the performances and their
implication for both VSI and Gravity VLTI instruments.
\end{abstract}


\keywords{Optical Interferometry, Integrated Optics}

\section{INTRODUCTION}

To pursue the development and increase the capabilities of the VLTI,
ESO selected three instrumental projects in phase A. The three new
instruments are : MATISSE, VSI, and GRAVITY. MATISSE is a mid-infrared
spectro-interferometer. VSI and GRAVITY are two near-infrared
instruments, both with the specificity of using integrated optics beam
combiners
\cite{1999A&AS..138..135M,1999A&AS..139..173B,2000ApOpt..39.2130H,
  2001A&A...376L..31B,2002A&A...390.1171L,2006A&A...450.1259L}. To
that end, new integrated optics (IO) beam combiners were developed.
The requirements for VSI were:
\begin{itemize}
\item Being able to work in the J, H and K band.
\item Being able to combine up to 6 telescopes.
\end{itemize}
on the other hand, GRAVITY had several criterium to meet:
\begin{itemize}
\item High sensitivity in the K band (more than 50\% throughput for the
  beam combiner as a whole).
\item Non-temporal modulation to allow long integration time exposures.
\end{itemize}

The basic concept for the combiner was an ABCD type-recombination
\cite{1977JOSA...67...81S}. Such a combiner was already presented by
M. Benisty et al. \cite{2006SPIE.6268E..73B}. Multi-axial 6 beam
combiners component are also investigated as a good alternative.  The
goal of this new development round is to: (i) prove the validity of
the technology for the K and J band, and (ii) improve the throughput
and therefore the sensitivity by testing different arrangement of the
waveguide paths. Several new combiners were therefore designed, and
were recently delivered to our laboratory.

\section{THE WAFER}

\begin{figure}
   \begin{center}
   \begin{tabular}{c}
   \includegraphics[height=13cm]{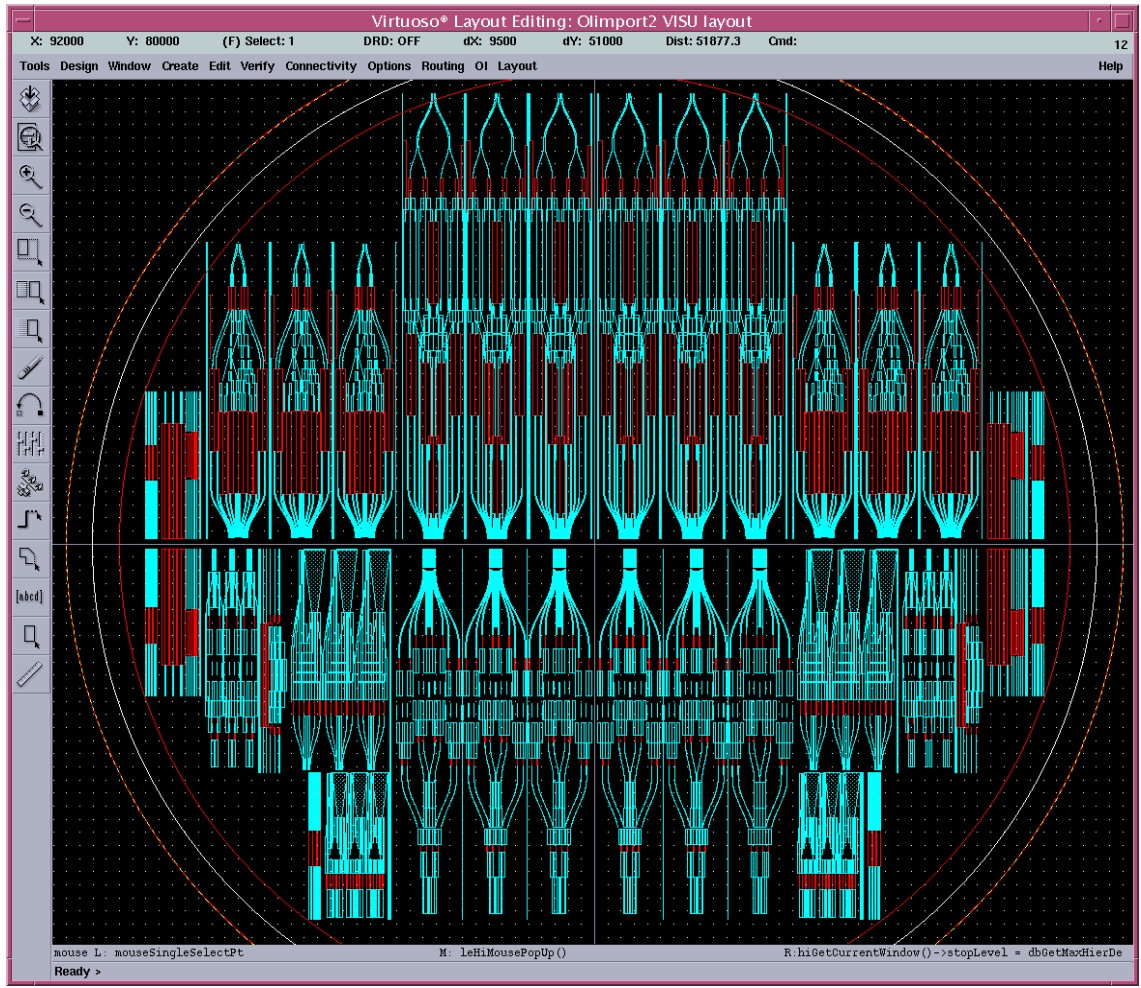}
   \end{tabular}
   \end{center}
   \caption[example] 
   { \label{fig:waf} 
Overview of the wafer guide tracing. On this $8\times8$\,inches wafer are 48 beam combiners.}
   \end{figure}

CEA/LETI technical processes use Silica on Silicum, and a lithographic
technique for waveguide tracing.  Figure~\ref{fig:waf} gives an
overview of the photomask which was used to transfer the waveguide
paths on the wafer. On this mask, there are 48 beam combiners. Each
type of combiner is duplicated on the right and the left of the
wafer. Each combiner also comes in three versions, corresponding to
guides of different size. For example, the H band combiners exist with
guides of 6.8$\,\mu$m 7$\,\mu$m and 7.2$\,\mu$m. The size of the wafer
is $8\times8$\,inches.


The new types of combiners are:
\begin{itemize}
\item A 4 telescopes ABCD beam combiner in the H band. The goal of this
  combiner is to test new achromatized phase shifts. This to (i)
  increase the contrast ratio at the output by compensating the chromatic 
  effects of the couplers, and (ii) increase the precision on the
  closure phases by having them constant over a large bandpass.
\item A second 4 telescopes ABCD beam combiner in H band. The size of
  the beam combiner is reduced by 30\% by using a combination of
  66/33\% and 50/50\% couplers instead of a tricoupler. The component
  is therefore not symmetric anymore, with an eventual effect on the
  closure phase.
\item Two 2 telescopes ABCD beam combiners to test transmission and response of the K
  and J band.
\item A 4 telescopes ABCD fringe tracking combiner in the K
  band. Since only pistons measurements are
  needed, such a combiner combine the flux using ``bootstrapping''
  approach (ie., each telescope is recombined with two other telescopes
  only).  The goal for this component is to validate fringe tracker
  algorithms with an ABCD combiner.
\item Two 6 telescopes multi-axial beam combiners, working in K and H bands
\end{itemize}

\section{CHARACTERIZATION I. TRANSMISSION}

Throughput was investigated by the means of two single mode fibers.
One is used to inject the light in the component, and the other to
convey the flux from the outputs of the component to a K band
monopixel detector. The flux recorded on all the outputs is then
summed, and normalized by the flux measured when putting the two
single mode fibers in contact with each other. Errors are of the order
of half a percent.

\begin{table}[h]
\caption{Transmission of the 2 telescopes ABCD beam combiner in the K band.} 
\label{tab:trans}
\begin{center}       
\begin{tabular}{|c|c|c|c|} 
\hline
\rule[-1ex]{0pt}{3.5ex}  Wavelength & Guide width & Input 1 & Input 2  \\
\hline
\hline
\rule[-1ex]{0pt}{3.5ex}  $2.20\,\mu$m & $9.8\,\mu$m & 52.0\,\% & 52.5\,\%\\
\hline 
\rule[-1ex]{0pt}{3.5ex}  $2.20\,\mu$m & $10.0\,\mu$m & 51.9\,\% & 53.4\,\%\\
\hline 
\rule[-1ex]{0pt}{3.5ex}  $2.20\,\mu$m & $10.2\,\mu$m & 60.0\,\% & 53.0\,\%\\
\hline 
\rule[-1ex]{0pt}{3.5ex}  $2.37\,\mu$m & $9.8\,\mu$m & 65.8\,\% & 66.2\,\%\\
\hline 
\rule[-1ex]{0pt}{3.5ex}  $2.37\,\mu$m & $10.0\,\mu$m & 68.4\,\% & 67.9\,\%\\
\hline 
\rule[-1ex]{0pt}{3.5ex}  $2.37\,\mu$m & $10.2\,\mu$m & 67.6\,\% & 68.9\,\%\\
\hline 
\end{tabular}
\end{center}
\end{table}

Tests were performed on the K band ABCD using two different SLED: one
at 2.2$\,\mu$m, and one at 2.37$\,\mu$m. Results are reported in
Table~\ref{tab:trans}. The 50\% transmission goal was achieved at both
wavelength. Optimization of the waveguide paths could eventually
further increase the transmission.

\section{CHARACTERIZATION II. COHERENT TRANSFER FUNCTION}

   \begin{figure}[h]
   \begin{center}
   \begin{tabular}{c}
   \includegraphics[height=4.5cm]{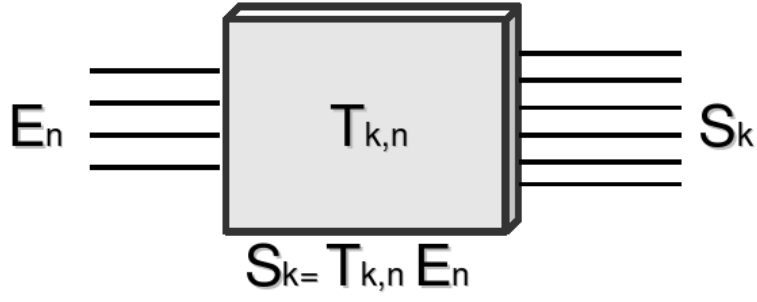}
   \end{tabular}
   \end{center}
   \caption[example] 
   { \label{fig:gen} Generalization of the transfer function of an
     integrated optics device. $n$ are inputs, $k$ outputs. $E_k$ and
     $S^k$ are respectively the entering and exiting complex electric
     field.}
   \end{figure}

   \begin{figure}
   \begin{center}
   \begin{tabular}{c}
   \includegraphics[height=4.5cm]{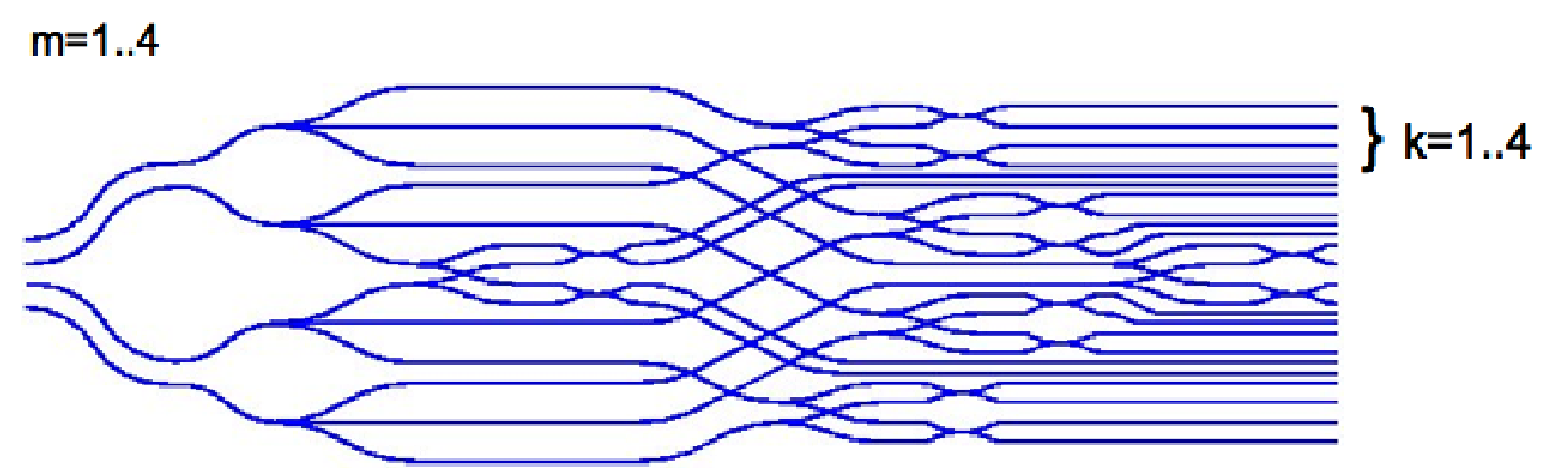}
   \end{tabular}
   \end{center}
   \caption[example] 
   { \label{fig:4t} }
   \end{figure} 

Figure~\ref{fig:gen} represents the generalized view of the transfer
function of an integrated optic component. $E_n$ is the complex
electric field entering the component via input $n$, and $S^k$ is the
resulting field on output number $k$. $T_{n}^k$ is a two dimensional
complex matrix linking $S^k$ to $E_n$.

A complete determination of the transfer function of the IO would
therefore be equivalent to the determination of matrix
$T_n^k$. However, the observable is not $S^k$ , but the intensity on
the output channels:
\begin{eqnarray}
|S^k|^2&=&\left|\sum_n T_n^k E_n\right|^2\\
&=&\Re \left[ \sum_n |T_n^k E_n|^2+2\sum_n\sum_{m>n}  T_n^kT_m^{k*}\ E_nE_m\right] \, .
\end{eqnarray}
However, this equation assume a fully coherent incoming beam, and no
loss of contrast due to, for example, chromaticity, polarization
etc... Hence, we introduced two extra terms: $V_{n,m}$ correspond to
the coherency of the incoming electric field (the complex
visibility). $C_{n,m}^k$ correspond to the level of which the IO
device conserve the coherence of the light:
\begin{equation}
|S^k|^2=\Re \left[ \sum_n |T_n^k E_n|^2+2\sum_n\sum_{m>n}  T_n^kT_m^{k*}C_{n,m}^k\ E_nE_m^*V_{n,m}\right] \, .
\end{equation}

Equation (3) can be rewritten with a matrix product:
\begin{equation}
\left(
\begin {array}{c}
|S^1|^2\\
\vdots \\
|S^K|^2
\end {array}
\right)
=  
\Re \left[ 
V2PM
\cdot
\left(
\begin {array}{c}
|E_1|^2\\
\vdots \\
|E_N|^2\\
E_{1}E_{2}^*V_{1,2}\\
\vdots\\
E_{N-1}E_{N}^*V_{N-1,N}\\
\end {array}
\right)
\right]
\end{equation}
where $N$ is the total number of inputs, and $K$ the number of
outputs. The $V2PM$ matrix is then equal to:
\begin{equation}
V2PM
=   
\left(
\begin {array}{cccccc}
|T_1^1|^2&\cdots&|T_N^1|^2& 2T_1^1T_2^{1*}C_{1,2}^1&\cdots&2T_{N-1}^1T_N^{1*}C_{N-1,N}^1\\
\vdots & & \vdots&\vdots & & \vdots\\
|T_1^K|^2&\cdots&|T_N^K|^2& 2T_1^KT_2^{K*}C_{1,2}^K&\cdots&2T_{N-1}^KT_N^{K*}C_{N-1,N}^K
\end {array}
\right)
\end{equation}
The name $V2PM$ was adopted in accordance with previous data reduction done on multi-axial interferometers\cite{2006MNRAS.368.1159T}. It stands for {``Visibilities to Pixels Matrix''}.

Characterization of the components will consist in the determination of
the V2PM. Providing a well engineered OI, the V22M shall be
invertible, allowing a robust determination of the photometry channels
($|E_n|^2$), as well as the contrasts and phases ($E_nE_m^*V_{n,m}$). 

We tested the method on the OI combiner already discussed by Benisty
et al.\cite{2006SPIE.6268E..73B}. The schematic of the combiner is
reproduced figure~\ref{fig:4t}. In the ideal case, the V2PM would
wrote :
\begin{equation}
V2PM
=    
\left(
\begin {array}{cccccccccc}
\op & \op & 0 & 0 & \oa & 0 & 0 & 0 & 0 & 0  \\
\op & \op & 0 & 0 & \ob & 0 & 0 & 0 & 0 & 0 \\
\op & \op & 0 & 0 & \oc & 0 & 0 & 0 & 0 & 0 \\
\op & \op & 0 & 0 & \od & 0 & 0 & 0 & 0 & 0 \\
 0 &\op & \op & 0 & 0 & \oa & 0 & 0 & 0 & 0  \\
 0 &\op & \op & 0 & 0 & \ob & 0 & 0 & 0 & 0  \\
\op &  0 &\op & 0 & 0 & 0 & 0 & 0 & 0 & \oa  \\
\op &  0 &\op & 0 & 0 & 0 & 0 & 0 & 0 & \ob  \\
\op & 0 & \op & 0 & 0 & 0 & 0 & 0 & 0 & \oc  \\
\op & 0 & \op & 0 & 0 & 0 & 0 & 0 & 0 & \od  \\
\op & 0 & 0 & \op & 0 & 0 & 0 & \oa & 0 & 0 \\
\op & 0 & 0 & \op & 0 & 0 & 0 &  \ob & 0 & 0  \\
\op & 0 & 0 & \op & 0 & 0 & 0 &  \oc & 0 & 0  \\
\op & 0 & 0 & \op & 0 & 0 & 0 &  \od & 0 & 0  \\
 0 &\op & 0 & \op & 0 & 0 & \oa & 0 & 0 & 0  \\
 0 &\op & 0 & \op & 0 & 0 & \ob & 0 & 0 & 0  \\
 0 &\op & 0 & \op & 0 & 0 & \oc & 0 & 0 & 0   \\
 0 &\op & 0 & \op & 0 & 0 & \od & 0 & 0 & 0   \\
 0 &\op & \op & 0 & 0 & \oc & 0 & 0 & 0 & 0  \\
 0 &\op & \op & 0 & 0 & \od & 0 & 0 & 0 & 0  \\
 0 & 0 &\op & \op & 0 & 0 & 0 & 0 & \oa & 0  \\
 0 & 0 &\op & \op & 0 & 0 & 0 & 0 & \ob & 0  \\
 0 & 0 &\op & \op & 0 & 0 & 0 & 0 & \oc & 0  \\
 0 & 0 &\op & \op & 0 & 0 & 0 & 0 & \od & 0  
\end {array}
\right)
\end{equation} 

\begingroup
\everymath{\scriptstyle}

There are several ways to obtained the matrix of ``real''combiner. We
used a method which determine the matrix column by column. The first
four columns correspond to the real transmission of each one of the
entrance separately. To determine these four columns, we inject light
into the entrance guides one after the other.  For the six
following column, light is injected into two entrances only, and
optical path modulated to reveals the phase shifts. 

 On one of our 4T
ABCD H band beam combiner, we obtained the following matrix:
\begin{equation}{
V2PM
=  
\left(
\begin {array}{rrrrrrrrrr}
0.085 & 0.039 & 0.001 & 0.001 & 0.112e^{0.00i} & 0.000e^{-2.02i} & 0.000e^{0.21i}& 0.001e^{-2.04i} & 0.000e^{2.79i} & 0.000e^{2.51i} \\ 
0.037 & 0.065 & 0.001 & 0.001 & 0.099e^{3.13i} & 0.001e^{1.25i} & 0.000e^{-1.65i}& 0.000e^{-0.52i} & 0.000e^{0.55i} & 0.000e^{-2.79i} \\ 
0.045 & 0.022 & 0.001 & 0.001 & 0.063e^{1.25i} & 0.000e^{-0.80i} & 0.000e^{-0.36i}& 0.000e^{1.24i} & 0.000e^{0.95i} & 0.000e^{-2.46i} \\ 
0.041 & 0.074 & 0.001 & 0.001 & 0.108e^{-1.88i} & 0.001e^{0.53i} & 0.000e^{2.82i}& 0.000e^{1.84i} & 0.000e^{-0.54i} & 0.001e^{2.28i} \\ 
0.002 & 0.067 & 0.054 & 0.002 & 0.002e^{2.27i} & 0.116e^{-0.00i} & 0.001e^{-3.11i}& 0.001e^{-2.01i} & 0.002e^{2.63i} & 0.001e^{-0.43i} \\ 
0.003 & 0.051 & 0.077 & 0.003 & 0.002e^{-1.03i} & 0.121e^{3.14i} & 0.001e^{2.40i}& 0.001e^{0.71i} & 0.000e^{1.28i} & 0.003e^{3.08i} \\ 
0.206 & 0.000 & 0.089 & 0.002 & 0.001e^{-0.09i} & 0.002e^{0.43i} & 0.000e^{-2.03i}& 0.002e^{-0.04i} & 0.001e^{2.79i} & 0.261e^{0.00i} \\ 
0.098 & 0.002 & 0.173 & 0.001 & 0.000e^{-2.84i} & 0.001e^{1.24i} & 0.000e^{2.99i}& 0.001e^{0.31i} & 0.001e^{0.19i} & 0.256e^{3.14i} \\ 
0.192 & 0.002 & 0.095 & 0.002 & 0.001e^{0.29i} & 0.002e^{2.23i} & 0.000e^{-2.10i}& 0.002e^{-2.92i} & 0.003e^{-0.99i} & 0.263e^{0.97i} \\ 
0.078 & 0.002 & 0.146 & 0.002 & 0.001e^{-1.19i} & 0.001e^{2.69i} & 0.000e^{2.71i}& 0.002e^{-0.09i} & 0.002e^{-1.57i} & 0.213e^{-2.16i} \\ 
0.093 & 0.003 & 0.000 & 0.035 & 0.001e^{-1.08i} & 0.001e^{2.44i} & 0.002e^{1.64i}& 0.107e^{-0.00i} & 0.001e^{-2.99i} & 0.004e^{0.88i} \\ 
0.044 & 0.004 & 0.004 & 0.101 & 0.001e^{2.30i} & 0.002e^{-2.90i} & 0.001e^{-1.75i}& 0.127e^{3.13i} & 0.002e^{-1.89i} & 0.001e^{1.40i} \\ 
0.037 & 0.003 & 0.003 & 0.014 & 0.001e^{2.59i} & 0.001e^{1.32i} & 0.000e^{1.36i}& 0.042e^{0.94i} & 0.001e^{-2.05i} & 0.001e^{2.74i} \\ 
0.020 & 0.002 & 0.002 & 0.043 & 0.001e^{-2.59i} & 0.000e^{2.44i} & 0.001e^{-0.86i}& 0.054e^{-2.15i} & 0.000e^{2.10i} & 0.001e^{-1.80i} \\ 
0.002 & 0.178 & 0.003 & 0.095 & 0.001e^{0.07i} & 0.001e^{-2.40i} & 0.237e^{0.00i}& 0.000e^{-1.37i} & 0.001e^{3.12i} & 0.001e^{-2.57i} \\ 
0.002 & 0.094 & 0.002 & 0.188 & 0.001e^{2.41i} & 0.001e^{1.62i} & 0.251e^{3.13i}& 0.001e^{-1.89i} & 0.001e^{-0.44i} & 0.000e^{-0.85i} \\ 
0.002 & 0.173 & 0.002 & 0.097 & 0.001e^{0.12i} & 0.001e^{0.75i} & 0.246e^{1.20i}& 0.001e^{-0.27i} & 0.001e^{-0.81i} & 0.001e^{1.86i} \\ 
0.002 & 0.089 & 0.002 & 0.187 & 0.000e^{-0.41i} & 0.000e^{-0.41i} & 0.245e^{-1.94i}& 0.001e^{-2.21i} & 0.000e^{-1.93i} & 0.001e^{-2.92i} \\ 
0.003 & 0.075 & 0.049 & 0.001 & 0.002e^{-2.88i} & 0.113e^{1.35i} & 0.003e^{1.60i}& 0.002e^{-1.19i} & 0.000e^{2.80i} & 0.002e^{-1.19i} \\ 
0.002 & 0.051 & 0.071 & 0.003 & 0.002e^{-1.16i} & 0.112e^{-1.77i} & 0.004e^{-2.65i}& 0.001e^{2.21i} & 0.004e^{1.99i} & 0.002e^{2.57i} \\ 
0.002 & 0.001 & 0.083 & 0.041 & 0.001e^{1.30i} & 0.001e^{1.68i} & 0.000e^{-2.87i}& 0.000e^{-0.59i} & 0.108e^{0.00i} & 0.001e^{-2.69i} \\ 
0.001 & 0.002 & 0.041 & 0.079 & 0.001e^{0.56i} & 0.001e^{1.52i} & 0.001e^{1.93i}& 0.001e^{2.17i} & 0.109e^{-3.14i} & 0.000e^{-1.67i} \\ 
0.001 & 0.001 & 0.068 & 0.036 & 0.000e^{-2.51i} & 0.000e^{2.92i} & 0.000e^{1.81i}& 0.001e^{-1.47i} & 0.094e^{1.26i} & 0.000e^{2.85i} \\ 
0.001 & 0.001 & 0.033 & 0.066 & 0.000e^{-2.86i} & 0.000e^{-0.14i} & 0.000e^{-2.12i}& 0.000e^{-0.01i} & 0.090e^{-1.86i} & 0.000e^{2.40i} \\ 
\end {array}
\right)}
\end{equation} 
\endgroup

The differences between matrix (5) and (6) are due several
instrumental effects. These effects
are revealed by :
\begin{itemize}
\item Missing zero values in the first four columns. This is due to
  crosstalk, consequence of some flux going from one guide to another
  because of, for example, intersections. In column 1, the crosstalk
  is of the order of 1\%.
\item Values different from $\op$ in the first four columns. This due to a transmission not perfectly equilibrated between the different channels after a coupler or a tricoupler.
\item Phase shifts different than $[0,\pi,\pi/2,3\pi/2]$ for the four main outputs of each telescope pairs. This is a problem due to the engineering of the phase shift. 
\end{itemize}

\section{CONCLUSION}

OI circuits presented in Figure~\ref{fig:waf} have recently been
delivered to our laboratory. We have shown that the transmission of
the components is adapted for astronomical observations in the K
band. Optimization of the throughput is nevertheless still under
investigation, with possibilities offered by using new couplers,
 optical paths, and/or fiber to combiner injection methods.

We also developed the tools to fully characterize the complex transfer
function of the new components, to have rigorous comparison between
the different combiner engineered. The knowledge of the $V2PM$ matrix
should also allow signal to noise estimation, to provide sound and
practical way to choose the best combiner for the VLTI.

\acknowledgments     
 
This project is funded by the French National Research Agency (ANR) –-- project 2G-VLTI.


\bibliography{report}   
\bibliographystyle{spiebib}   

\end{document}